# Casimir forces on a silicon micromechanical chip

J. Zou,[1,2] Z. Marcet,[1,2] A. W. Rodriguez,[3,4] M. T. H. Reid,[5] A. P. McCauley,[6] I. I. Kravchenko,[7] T. Lu,[2] Y. Bao,[1] S. G. Johnson,[4] H. B. Chan[2*]

[1]Department of Physics, University of Florida, Gainesville, Florida 32611, USA,

[2]Department of Physics, the Hong Kong University of Science and Technology, Clear Water Bay, Kowloon, Hong Kong, China,

[3]School of Engineering and Applied Sciences, Harvard University, Cambridge, Massachusetts 02138, USA,

[4]Department of Mathematics, Massachusetts Institute of Technology, Cambridge, Massachusetts 02139, USA,

[5]Research Laboratory of Electronics, Massachusetts Institute of Technology, Cambridge, Massachusetts 02139, USA,

[6]Department of Physics, Massachusetts Institute of Technology, Cambridge, Massachusetts 02139, USA,

[7]Center for Nanophase Materials Sciences, Oak Ridge National Laboratory, Oak Ridge, Tennessee 37830, USA;

*To whom correspondence should be addressed. Email: hochan@ust.hk

Abstract

Quantum fluctuations give rise to van der Waals and Casimir forces that dominate the interaction between electrically neutral objects at sub-micron separations. Under the trend of miniaturization, such quantum electrodynamical effects are expected to play an important role in micro- and nano-mechanical devices. Nevertheless, utilization of Casimir forces on the chip level remains a major challenge because all experiments so far require an external object to be manually positioned close to the mechanical element. Here, by integrating a force-sensing micromechanical beam and an electrostatic actuator on a single chip, we demonstrate the Casimir effect between two micromachined silicon components on the same substrate. A high degree of parallelism between the two near-planar interacting surfaces can be achieved because they are defined in a single lithographic step. Apart from providing a compact platform for Casimir force measurements, this scheme also opens the possibility of tailoring the Casimir force using lithographically defined components of non-conventional shapes.

# Introduction

The Casimir force[1] refers to the interaction between neutral objects that originates from the boundary conditions imposed on the zero point quantum fluctuations of the electromagnetic field. Besides fundamental interest, the Casimir force is also of technical importance in micro- and nano-mechanical systems, because it is believed to play a role in stiction, which stands for the permanent adhesion of movable components to nearby fixed surfaces[2,3]. Since the first precise measurement of the Casimir force more than a decade ago[4], significant progress has been made towards the engineering and control of Casimir forces[5,6]. In particular, optical properties and geometry effects provide powerful tools for tailoring the Casimir force. For example, repulsive Casimir forces in the retarded limit were demonstrated in experiments involving fluids[7,8]. Metamaterials, with their remarkable optical properties, have been considered as candidates for controlling the Casimir force[9,10], but initial suggestions that the sign of the force in vacuum could be altered by metamaterials turned out to be unrealistic[11]. With regard to the non-trivial dependence of the Casimir force on the shape of the bodies[12-15], experiments involving nanostructured surfaces have demonstrated the non-pairwise additive nature of the Casimir force[16,17]. Recent experiments have also measured the corrections to the Casimir force that arise from the presence of thermal fluctuations, in addition to quantum fluctuations[18].

The possibility for the Casimir force to play a role in micromechanical systems has been a major driver behind experimental research. However, demonstration of Casimir forces between micromachined surfaces in a single micromechanical chip has remained elusive, because standard experimental schemes require an external object to be manually positioned close to either cantilevers or torsional balances[4,8,19-22]. Bulky micropositioners and piezoelectric actuators are required to control the separation between the two interacting bodies. Such arrangements have hindered progress in the on-chip exploitation of the Casimir force. Conventional experimental setups also face a number of other challenges. For instance, maintaining the parallelism of two flat surfaces at small distances has proven to be difficult. As a result, in most experiments one of the two objects is chosen to be spherical. So far, there has only been one experiment that measured the Casimir force between two parallel plates[19]. The alignment becomes even more challenging for nanostructured surfaces. In fact, when corrugations are present on both surfaces, it is necessary to use an in-situ imprint technique such that the patterns are automatically aligned after fabrication[17]. Another major difficulty in measuring the Casimir force at room temperature is the long-term drift in the distance between the surfaces: since the distance from the two interacting elements to their common point of support typically measures at least a few centimeters, temperature fluctuations lead to uncontrollable distance variations, limiting the duration of measurement and hence the force resolution.

We demonstrate that the Casimir force can be the dominant interaction between single-crystalline silicon components on a semiconductor chip, in the absence of external objects. Both the force sensing element and the actuator that controls the distance are integrated on the same substrate. They are created by dry etching that produces sidewalls that are largely vertical. No external alignment of the interacting bodies is necessary because they are defined in a single lithographic step. The use of electron beam lithography ensures a high degree of parallelism between the interacting bodies in a near-planar geometry. Another advantage is that the distance of the interacting elements to their common support is reduced to ~ 70 µm, about a factor of 1000 smaller than conventional experiments. The improved mechanical stability minimizes long-term drifts in the gap between the interacting objects ($<10^{-5}$ nm s$^{-1}$). Furthermore, this scheme also allows tailoring of the Casimir force in the future using lithographically defined components of non-conventional shapes.

## Results

### Device and Measurement scheme

Figures 1a and 1b show a simplified schematic of the structure and a scanning electron micrograph of sample A, respectively. The element for sensing the force is a doubly-clamped silicon beam that is 100 µm long and ~ 1.42 µm wide, depicted in the top parts of Figs. 1a and 1b. The silicon is p-doped with a high carrier concentration of $7.0 \times 10^{18}$ cm$^{-3}$. Figure 1c zooms in on the micro-beam. A silicon electrode of width 2.80 µm is positioned close to the beam, as shown in the lower part of Fig. 1c. The beam and the electrode have the same thickness (2.65 µm) and distance to the substrate (2 µm) (see Methods, Supplementary Figure S1, and Supplementary Methods for fabrication details). Electrostatic and/or Casimir forces are exerted by the electrode on the beam depending on the voltage $V_e$ between them. The gap between the beam and the electrode is created by deep reactive ion etching while the beam and the electrode are protected with an etch mask. With the etch mask defined by electron beam lithography, a high degree of parallelism is ensured between the beam and the electrode, without any need for manual alignment prior to force measurement. By viewing the beam and the electrode from the top with a scanning electron microscope, their separation is found to remain constant to within ~15 nm along the entire length of the beam, yielding an upper bound of 150 μrad for the angle between the lithographic patterns of the beam and the electrode.

The electrode is attached to a comb actuator so that it can be controllably moved along the y direction, reducing the separation $d$ between the electrode and the beam from an initial distance of $d_0 = 1.92$ μm (measured with a scanning electron microscope) down to ~ 260 nm while maintaining parallelism (see Methods, Supplementary Figures, and Supplementary Methods). Figure 1d shows a close-up of part of the comb actuator. The comb actuator consists of a set of movable comb fingers supported by four serpentine springs (Fig. 1e), one at each corner of the structure (Fig. 1a). A second set of comb fingers (the solid structures in Fig. 1d with no etch holes) is fixed to the substrate on one end. When a voltage $V_{comb}$ is applied to the fixed comb relative to the movable comb (in the experiment, a negative $V_{comb}$ is used), an electrostatic force parallel to the substrate is generated. The movable combs are displaced towards the beam until the restoring force from the four springs balances the electrostatic force. As a result, the separation $d$ is reduced as $|V_{comb}|$ increases (see Supplementary Methods and Supplementary Movies). It should be noted that the electrostatic force between the fixed and movable combs merely serves to set $d$. As explained below, the potential difference between the beam and the movable electrode $V_e$ can be controlled separately, independent of $V_{comb}$.

The suspended beam acts as a resonant force sensor. As shown in Fig. 1a, a small ac voltage ($V_{ac} = 5.7$ μV) is applied to one end of the beam, producing an ac current. In the presence of a 5 T magnetic field perpendicular to the substrate, the beam is subjected to a periodic Lorentz force. The frequency of the ac voltage $\omega_D$ is chosen such that the beam vibrates along the y direction in its fundamental mode. Vibrations of the beam in the magnetic field generate an induced electromotive force that is detected with a current amplifier. In Fig. 2a, the resonant frequency of the beam $\omega_R$ is measured to be $7.26185 \times 10^6$ rad/s. All measurements were performed at 4 K and $< 10^{-5}$ Torr. When the Casimir force and/or electrostatic forces are exerted on the beam, the resonant frequency $\omega_R$ decreases due to the spring softening effect, by an amount $\Delta\omega_R$ that is proportional to the force gradient $F'(d)$:

$$\Delta\omega_R = K F'(d), \tag{1}$$

where $K$ is a positive proportionality constant and $F'(d) < 0$.

**Calibration**

Similar to conventional experiments on Casimir forces, we also need to perform a calibration procedure by applying a voltage $V_e$ between the beam and the movable electrode to generate an electrostatic force $F_e$ between them. $F_e$ is proportional to $(V_e - V_0)^2$, where $V_0$ is the residual voltage. Figure 2b shows parabolic fits to $\Delta\omega_R$ versus $V_e$. Each parabola corresponds to $\Delta\omega_R$ recorded at a fixed $d$ that is set by $V_{comb}$. There are two contributions to $\Delta\omega_R$: the electrostatic part that depends quadratically on $V_e - V_0$ and a vertical offset that is independent of $V_e - V_0$. The latter becomes more negative as $d$ decreases. As described later, we will compare this vertical offset to the Casimir force gradient and remnant force gradients due to patch potentials. The electrostatic part will be used for force calibration.

The residual voltage $V_0$ is measured by identifying $V_e$ at which the maximum of the parabolic dependence of $\Delta\omega_R$ occurs. Figure 2c shows that $V_0$ is measured to be about -25 mV at small $d$. Over the full range of distances, $V_0$ changes by about 15 mV, comparable to previous experiments in the lens-plate[23] and sphere-plate[24] geometries.

Even though both the beam and the electrode are made of single-crystal silicon on the same wafer, the residual voltage $V_0$ is non-zero and shows distance dependence. Based on current experimental data, we cannot convincingly identify the origin of these effects. We suspect that the non-zero residual voltage is possibly due to solder contacts in the electrical leads at different temperatures. The distance dependence likely originates from adsorbed impurities and/or the etching profile (see Supplementary Methods) exposing patches of different crystal orientations at the sidewalls with non-uniform potentials.

In conventional Casimir force experiments, the extension of the piezoelectric element is either pre-calibrated or directly measured. At the same time, the initial distance between the two interacting surfaces is an unknown that needs to be determined by the application of electrostatic forces. In our experiment, distance calibration is performed using a slightly different procedure. Here, $d$ is given by:

$$d = d_0 - \alpha V_{comb}^2, \qquad (2)$$

where $d_0$ is the initial separation, $V_{comb}$ is the voltage of the fixed combs relative to the movable combs and $\alpha = \dfrac{1}{2k_{//}} \dfrac{dC_{comb}}{dy}$ is a proportionality constant to be determined by fitting. $k_{//}$ is the spring constant of the serpentine springs along y. $\dfrac{dC_{comb}}{dy}$, the spatial derivative of the capacitance between the fixed and the movable combs, remains almost constant as the movable combs are displaced. One main difference from previous experiments is that $d_0$ is not a fitting parameter. Instead, $d_0$ is accurately measured to be $1.92 \pm 0.015$ μm using a scanning electron microscope. The dependence of $F_e$ on the distance $d$ is calculated using finite element analysis with our device geometry. By fitting to the calculated electrostatic force gradient $F'_e(d)$ given by Eqs. (1) and (2) for six sets of data with $V_e$ ranging from $V_0 + 100$ mV to $V_0 + 150$ mV, we obtain $\alpha = 10.55 \pm 0.04$ nm V$^{-2}$ and $K = 5.38 \pm 0.10 \times 10^4$ rad m (s N)$^{-1}$. Figure 2d plots the electrostatic force gradient on the beam as a function of $d$ at $V_e = V_0 + 100$ mV, where $d$ is controlled by increasing $|V_{comb}|$ from 0 to 11.375 V according to Eq. (2). When $V_e$ is applied, the effective distance for the electrostatic force changes because of carrier depletion. However, due to the high carrier concentration ($7 \times 10^{18}$ cm$^{-3}$), this change is small (< 1 nm) and negligible compared to the uncertainty in $d$.

**Casimir Force Measurement and Calculations**

Next, we set $V_e = V_0(d)$ for each distance $d$ and measure the force gradient $F"_c$ between the beam and the electrode as a function of $d$ (Fig. 3a). The red line in Fig. 3a represents the theoretical values of the Casimir force calculated for silicon structures of such geometry, with no fitting parameters. The theoretical calculation involves a boundary-element method (BEM) discretization of the beam and substrate surfaces, combined with a recent fluctuating-surface-current formulation of the Casimir force between dielectric bodies that writes the full Casimir-energy path integral as a simple expression in the classical BEM interaction matrix[14,25]. It includes the contributions of the finite conductivity of silicon and the imperfect etching profiles on the sidewalls of the beam and the electrode (~ 88° from the substrate surface, see Supplementary Methods). Despite the imperfect agreement between measurement and theory, it is clear that the Casimir force becomes the dominant interaction between the beam and the movable electrode at small $d$. Unlike the sphere-plate configuration, the roughness on the sidewalls cannot be directly measured. From the top view micrograph of the beam and electrode, we determine the rms roughness of the edges to be 12 nm, mainly due to non-uniformity of the electron beam lithography. The roughness correction is estimated to be about 3% of the Casimir force at the closest distance[26]. We further note that the calculation assumes the lithographic patterns of the beam and the electrode to be parallel. Using an upper bound of 150 µrad for the angle between them, the calculated Casimir force increases by up to 1.1% at the closest distance[26].

One plausible explanation for the deviations involves residual electrostatic forces that are associated with patch potentials. Following the analysis of ref. 23, such residual forces display similar distance dependence as the applied electrostatic force, but varies as $[V_0(d) + V_1]^2 + V_{rms}^2$. The first term originates from the spatial variations in the contact potential that give rise to a distance dependent $V_0$ while the second term accounts for patches smaller than the effective interaction area. $V_1$ accounts for a constant offset of the averaged contact voltage at large distances while $V_{rms}$ is the rms value of the random patch voltages. Currently, our experiment lacks the sensitivity to resolve the force gradients at $d > 0.8$ µm. Therefore, unlike ref. 23, we cannot measure long range residual electrostatic forces and fit at distances where the Casimir force is negligible. Instead, we fit the measured force gradients to the calculated Casimir force gradient modified by residual electrostatic forces due to patch potentials (purple line in Fig. 3a). By setting $V_1 = -V_0 = 11.3$ mV at the largest $d$, we obtain a fitted value of $V_{rms} = 16.8$ mV.

## Discussion

The Casimir force gradient between a beam and an electrode with near-square cross sections has not been measured experimentally before. This configuration opens the possibility of testing a number of fundamental concepts. One important question is the validity of the proximity force approximation (PFA)[27]. The inset of Fig. 3a compares the

Casimir force gradient generated by the PFA to calculations of the exact geometry of the silicon structures. The ratio decreases with distance, reaching 56% at 6 µm. Secondly, this geometry could offer a direct demonstration of the non-pairwise nature of the Casimir force. The inset of Fig. 3a shows that the calculated Casimir force between the beam and the electrode depends on the presence of a third body, the substrate. When the substrate is removed, the Casimir force increases by ~ 14% at 6 µm. We note that the PFA yields the same force on the beam regardless of whether a substrate is present, as interactions with the third body are not taken into account. Recent theoretical analysis also predicted that the Casimir force between the beams exhibits a non-monotonic dependence on the distance to the substrate that cannot be explained by pairwise additive models of the force[28]. Since the Casimir force gradient at these distances is beyond the reach of the current setup, future experiments to reveal the aforementioned effects would require more sophisticated measurement circuitry or other detection schemes[29-31] to improve the sensitivity at large (> 2 µm) separations. Alternatively, beams and electrodes with smaller cross sections and/or smaller beam–substrate separations can be used to generate Casimir forces that are distinguishable from the PFA and pairwise additivity at smaller distances. Further progress will also require a better understanding of the force from the patches potentials, in order to suppress such forces or separate them from the Casimir force in the total measured force.

We have shown that the Casimir force can be the dominant interaction between silicon micromechanical components on an integrated micromechanical chip, in the absence of external objects. By creating the two interacting surfaces using nanofabrication, we circumvented the problem of aligning two planar surfaces. Our results represent the first step towards on-chip exploitation of the Casimir force. The compact and integrated actuation and detection platform provides new opportunities of tailoring the Casimir force using geometry effects. Structures are defined by electron beam lithography, making it possible to create surfaces with complex shapes[32] that are automatically aligned after fabrication. Furthermore, the design can be integrated with suspended planar waveguide elements so that classical optical gradient forces[31] and Casimir forces associated with virtual photons can be simultaneously utilized in these optomechanical systems. Nevertheless, much progress will be necessary, particularly in sample fabrication and characterization, to achieve the < 5% agreement between theory and measurement[18,21,22] that has been claimed in conventional Casimir force experiments between spheres and plates.

**Acknowledgements**

H.B.C., J.Z., Z.M. and Y.B. are supported by DOE No. DE-FG02-05ER46247 and NSF No. DMR-0645448. H.B.C. and T. L. are supported by Shun Hing Solid State Clusters Lab and HKUST 600511 from the Research Grants Council of Hong Kong SAR. S.G.J., A.W.R., and A.P.M are supported in part by DARPA under contract N66001-09-1-2070-DOD. M.T.H.R. was supported by the Singapore-MIT Alliance's Program in Computational Engineering. A portion of this research was conducted at the Center for Nanophase Materials Sciences, which is sponsored at Oak Ridge National Laboratory by the Office of Basic Energy Sciences, U.S. Department of Energy.

**Contributions**

J.Z. and H.B.C. conceived the experiment. J.Z., Z.M., I.K. and Y.B. designed, fabricated and characterized the devices. Z.M. and J.Z. conducted the measurements. Z.M., J.Z., T.L. and H.B.C. analyzed the data. A.W.R., M.T.H.R., A.P.M. and S.G.J. performed the theoretical calculations. H.B.C. supervised the experiment. All authors discussed the result and contributed to the writing.

**Competing financial interests**

The authors declare no competing financial interests.

**Corresponding author**

Correspondence and requests for materials should be addressed to H.B.C.

Figure Captions

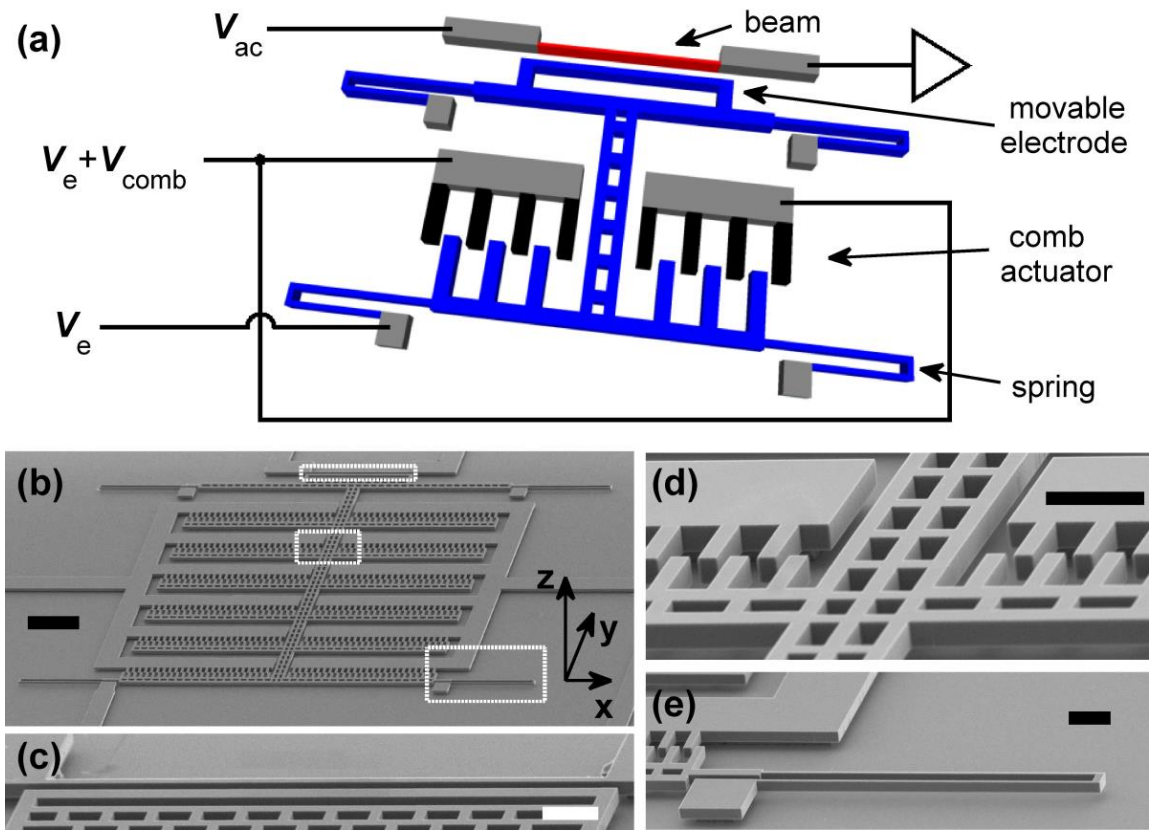

**Figure 1 | The set-up of the experiment and device. a**, A simplified schematic (not to scale) of the beam (red), movable electrode and comb actuator supported by four springs (blue), with electrical connections. The current amplifier provides a virtual ground to the right end of the beam. Suspended and anchored parts of the comb actuator are shown in blue and dark grey respectively. The separation $d$ between the beam and the movable electrode was controllably reduced so that the Casimir force can be detected. **b-e**, Scanning electron micrographs of the entire micromechanical structure (**b**) and close-ups of: the doubly clamped beam (**c**), the comb actuator (**d**) and the serpentine spring (**e**). The close-ups in (**c-e**) zoom into the top, middle, and bottom white dashed boxes in (**b**), respectively. The scale bars represent 50 µm in (**b**) and 10 µm in (**c-e**).

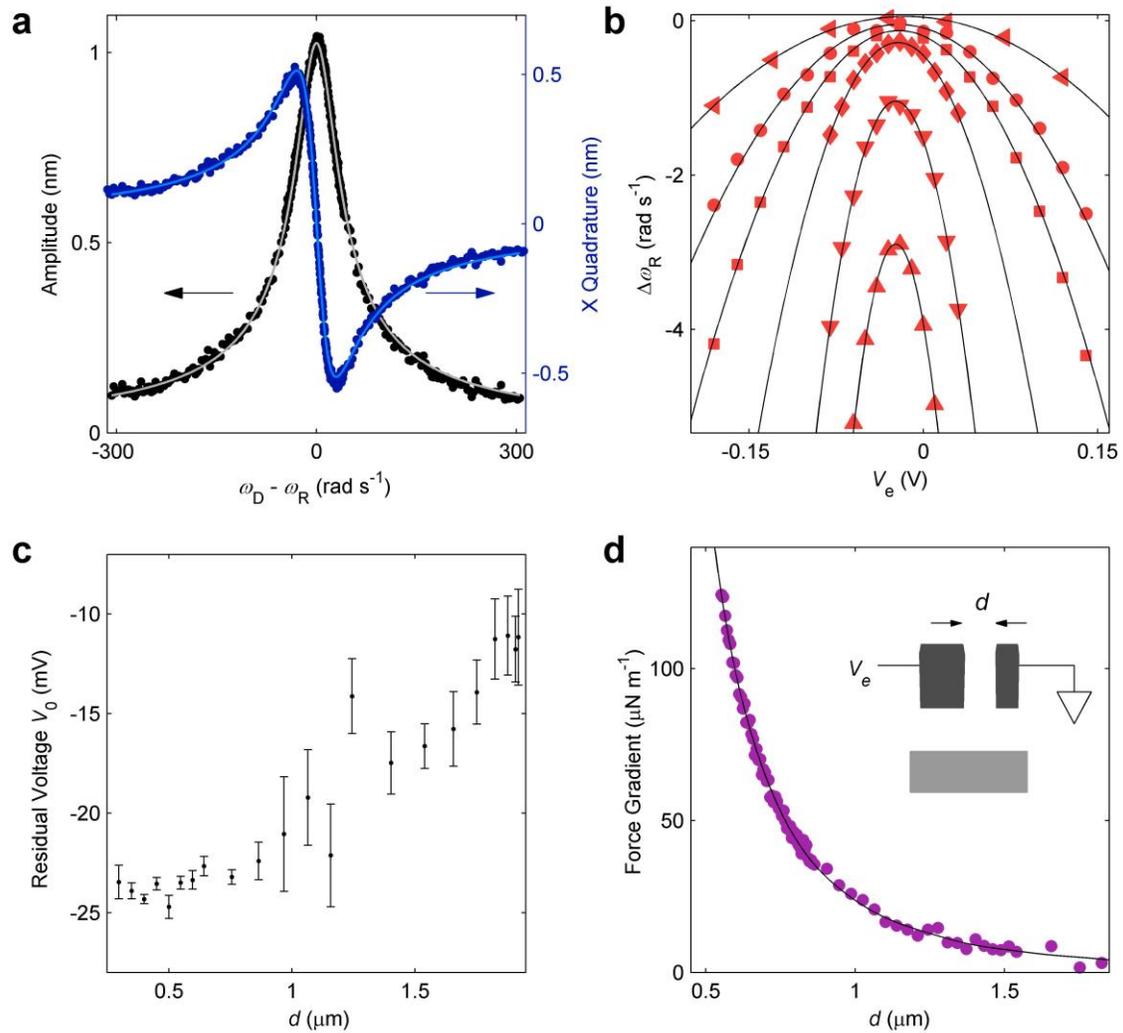

**Figure 2 | Calibration of the device using electrostatic force gradient. a**, Oscillation amplitude of the beam and its X quadrature that is in phase with the periodic driving force. Fitting to the driven underdamped oscillator model (lines) gives a damping coefficient of 30 rad s$^{-1}$. **b**, Measured frequency shift $\Delta\omega_R$ as a function of electrode voltage $V_e$, at $d$ = 1.403 µm, 1.065 µm, 865 nm, 643 nm, 450 nm and 349 nm, from top to bottom. **c**, Measured dependence of the residual voltage as a function of $d$. **d**, Measured electrostatic force gradient on the beam (circles) at $V_e = V_0 + 100$ mV. The line represents a fit to the values

calculated using finite element analysis. Inset: cross-sectional schematic of the beam, electrode and substrate with dc electrical connections. Error bars represent plus and minus one standard error.

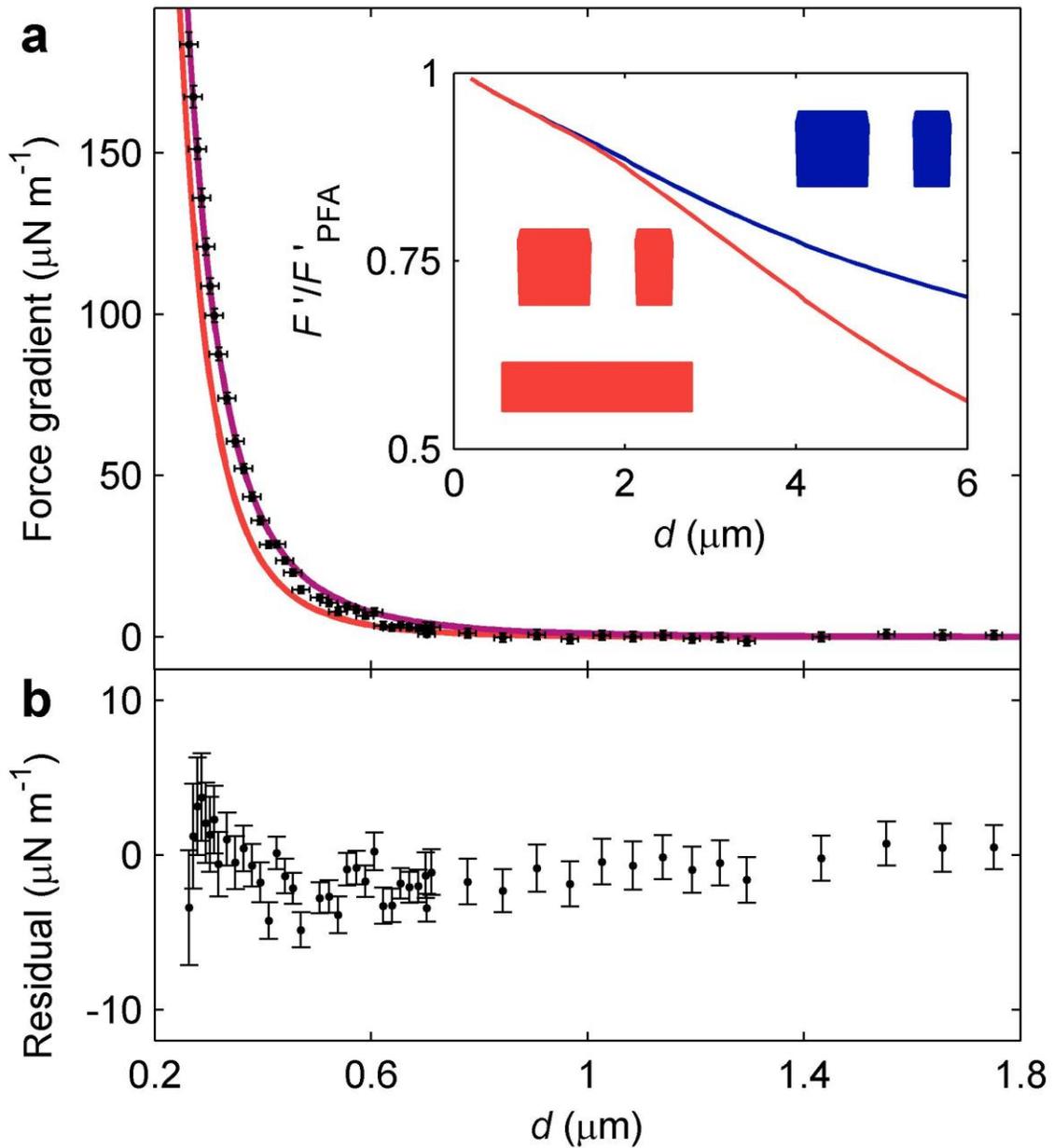

**Figure 3 | Measured force gradient $F'_c$ between the beam and the movable electrode as a function of separation $d$ after compensating for the residual**

**voltage. a,** The red line represents the calculated Casimir force gradients between an electrode and a beam made of silicon. The purple line includes possible contributions from patch potentials. Inset: The ratios of the calculated Casimir force between the beam and the electrode to the forces given by the PFA are plotted as the red (with substrate) and blue lines (without substrate). **b**, Deviations of the measured force gradient from the purple line in **a**. Error bars represent plus and minus one standard error.

**METHODS**

**Device fabrication and preparation.** The device is fabricated using a silicon-on-insulator (SOI) wafer with a 2.65-um silicon device layer and a 2.0-um buried oxide layer, so that the beam, electrode and comb actuator have identical thickness and distance from the substrate. The gap between the beam and the plate is created by deep reactive ion etching of the silicon device-layer while the beam and the electrode are protected with an etch mask defined by electron-beam lithography. After removing the etch mask and the underlying sacrificial silicon oxide layer with hydrofluoric acid (HF), the beam is free to vibrate in response to external forces (see Supplementary Methods).

Both the device layer and the handle wafer are p-doped with boron. The resistivity and carrier concentration of the device layer are measured to be 0.011 $\Omega$ cm and $7.0 \times 10^{18}$ cm$^{-3}$ respectively at 4 K by the van der Pauw method. For the handle wafer, the resistivity is measured to be 21.5 $\Omega$ cm at room temperature. We find that it becomes effectively insulating at 4 K.

To avoid charges being trapped on the surfaces of the beam and the movable electrode, HF is used to remove the native oxide on silicon right before force measurement. HF also passivates the surfaces to temporarily prevent the re-formation of oxide at ambient pressure[16]. The chip is then immediately loaded into a sealed probe that is evacuated and then lowered into a cryostat.

**Measurement of the force gradient on the beam.** The shift $\Delta\omega_R$ is measured by maintaining the ac voltage at a fixed frequency and recording the change in the X quadrature. Provided that $\omega_R$ remains close to the frequency of the ac voltage, small changes in the X quadrature is proportional to $\Delta\omega_R$, which is in turn proportional to $F'(d)$.

$\Delta\omega_R$ can be inferred using the change in the X quadrature and the slope of the X quadrature response (the blue curve in Fig. 2a). At the largest measured $\Delta\omega_R$, we estimate the deviation from a linear relationship between the X quadrature and $\Delta\omega_R$ to be less than 2.5%, which is smaller than our measurement uncertainty. The oscillation amplitude of the beam is kept small ($< 1.4$ nm $\ll d$). We calibrate the oscillation amplitude of the beam by measuring the onset of nonlinear mechanical bistability. Static deflection of the beam in response to the electrostatic and Casimir forces from the movable electrode is less than 2 pm, negligible compared to $d$. We do not resolve any change in the width of the amplitude response (Fig. 2a) of the beam with distance.

**Geometry characterization.** As described in the main text, the nanofabrication process produces two interacting surfaces that are automatically aligned and almost parallel to each other. Deviations from the ideal parallel-plane geometry may occur due to imperfections in the processing steps. For example, drift and misalignment of the electron-beam writer may produce a lithographic pattern that deviates from perfect parallelism. Using scanning electron microscopy, we analyzed the close-up top views of the gap close to the two ends of the beam and at different locations of the beam. Within the measurement uncertainty, we did not observe any change of the gap across the beam. We estimate that the upper bound for the angle between the lithographic patterns of the beam and the electrode is 150 μrad (corresponding to a gap changing 15 nm over 100 μm). Using the above parameters, the force increases by less than 1.1% at the smallest separation.

For a structure with dimensions exactly equal to the nominal values in the electron beam lithographic pattern, the angle between the beam and the movable electrode remains unchanged as *d* decreases. In the actual device, however, non-uniformities in exposure and etching might produce small asymmetries in the structure. To estimate such effects, we consider the case in which the two springs on the left side is wider than the ones on the right side by 50 nm. This value is a conservative estimate because it significantly exceeds the resolution of the electron beam writer. Numerical calculations show that at the largest $|V_{comb}|$, the movable electrode is closer to the beam by less than 4 nm on the right side than the left side. This value is smaller than the roughness of the sidewalls of the beam and the movable electrode and leads to a negligible tilting angle ~ 40 μrad. Our calculation indicates that it increases the Casimir force by less than 0.1% at the closest distance.

The cross sectional profiles of the beam and the electrode were also analyzed, by cleaving a second sample that is fabricated together with the sample in which the Casimir force was measured (Supplementary Figure S3). The slightly convex slope near the top of the cross section arises due to the periphery of the aluminum etch mask being consumed in the dry etching process. For the rest of the sidewall, the DRIE etching recipe produces a concave profile at an angle of ~ 88° to the top surface. This geometry was used in the BEM numerical calculation and the PFA. More details are discussed in Supplementary Methods.

**Calculations of the Casimir force.** The theoretical calculation (red line in Fig. 3a) neglects the finite length of the beams. The geometry reduces to a 2D problem in the cross sections (integrated over the longitudinal wavevector), and each object's surface is discretized into set of line segments described by "rooftop" basis functions[25]. We found that a discretization of approximately 3200 total points for all surfaces and a substrate truncated to 1 µm was sufficient to obtain convergence to 1% accuracy.

In the calculations, the dielectric function of silicon $\varepsilon(i\xi)$ is obtained using the Kramers-Kronig relations with the tabulated optical constants for frequency $\xi > 5000$ rad s$^{-1}$. For $\xi < 5000$ rad s$^{-1}$, $\varepsilon(i\xi)$ is given by:

$$\varepsilon(i\xi) = 1.0035 + (11.87 - 1.0035)/(1 + \xi^2/\omega_0^2) + \omega_p^2/[\xi(\xi + \Gamma)], \qquad (3)$$

where $\omega_0 = 6.6 \times 10^{15}$ rad s$^{-1}$, $\omega_p = 2.5562 \times 10^{14}$ rad s$^{-1}$ and $\Gamma = 6.364 \times 10^{13}$ rad s$^{-1}$.

In estimating the roughness correction, the force between two flat plates with stochastic roughness is first calculated using Eq. 16 in ref. 26 as a function of distance. Then the PFA is used to generate the force for our geometry.

# Supplementary Information for

# Casimir forces on a silicon micromechanical chip

**Supplementary Figures**

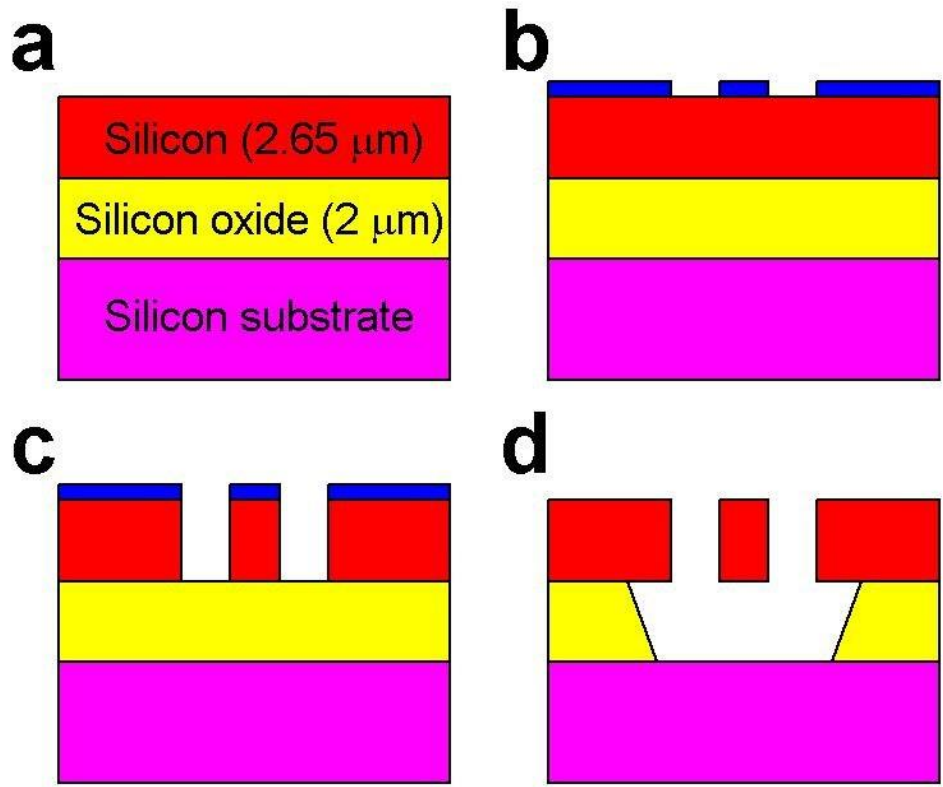

**Supplementary Figure S1 | The fabrication procedure of the device**. **a**, A silicon-on-insulator wafer is used to fabricate the device. **b**, The aluminum etch mask is created by lift-off on a resist pattern written by electron beam lithography. **c**, DRIE removes the silicon in the regions not protected by aluminum. The etch stops at the oxide layer. **d**, An isotropic wet etch by HF selectively removes the silicon oxide, producing undercuts of the top structure by ~ 2.7 μm. The middle piece of silicon is suspended while the other two pieces are anchored to the substrate through the remaining oxide. (Not to scale.)

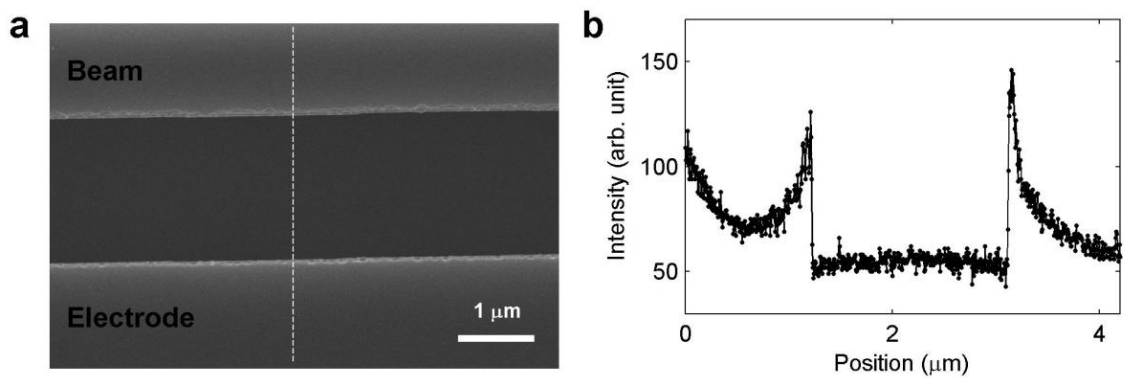

**Supplementary Figure S2 | The analysis of the initial distance $d_0$. a**, Scanning electron micrograph of the top view of the beam and the movable electrode. **b**, Intensity vs. distance along the dotted line in **a**. Adjacent pixels are separated by 5 nm. The zero in position refers to the top of the dotted line in **a**.

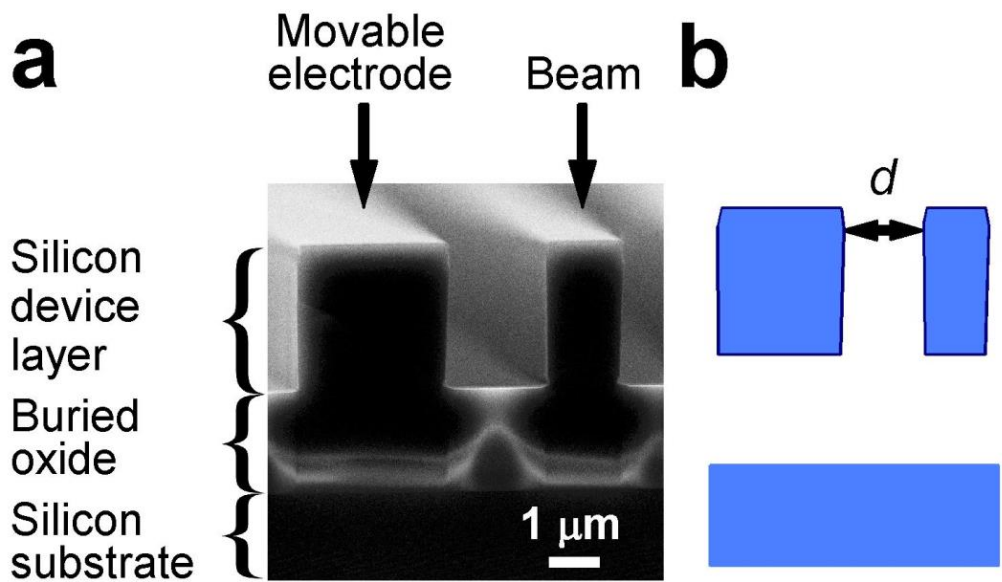

**Supplementary Figure S3 | The geometry of the interacting components**. **a**, Scanning electron micrograph of the cross section of sample B fabricated simultaneously with sample A in which the Casimir force was measured. The buried oxide layer has not been etched away. **b**, The exact cross-sectional geometry used in the calculation of electrostatic and Casimir forces.

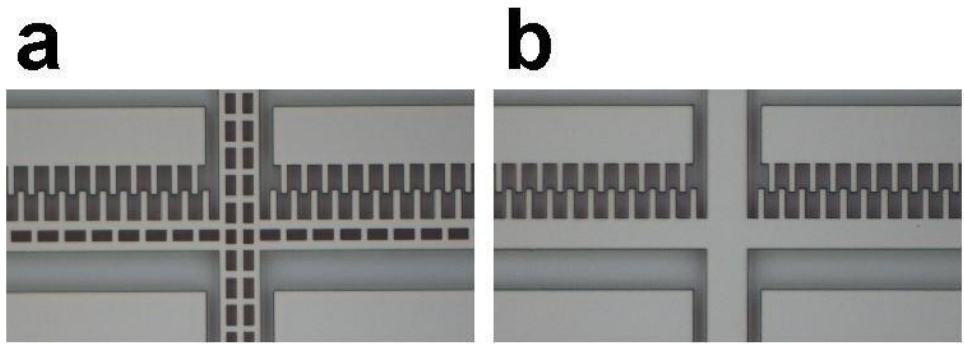

**Supplementary Figure S4 | Optical micrographs of samples C and D**. **a,** Optical micrographs of sample C in which the comb drives have etching holes. They become suspended after the underlying oxide is removed by hydrofluoric acid. **b,** sample D with no etch holes on the comb drives. The latter is used to verify that $V_{comb}$ has no measurable effects on the beam if the electrode is immobile.

**Supplementary Methods**

1. Sample Design and Fabrication

Supplementary Figure S1 shows the process for fabricating the device. Electron beam lithography with proximity correction is used to produce the resist pattern on the SOI wafer. An aluminum film of thickness ~ 90 nm is evaporated onto the sample followed by lift-off. The smallest achievable characteristic dimension is ~ 20 nm. Using the aluminum as an etch mask, the unprotected silicon is then removed by deep reactive ion etching (DRIE). A continuous etch and passivate recipe is chosen to produce sidewalls with no undulations. Afterwards, an oxygen plasma etch removes the residual hydrocarbon generated during DRIE. The chip is then placed in hydrofluoric acid to etch the buried oxide to release the structures that are designed to be movable. The duration of this isotropic wet etch is chosen so that in addition to removing the oxide in regions that are exposed by DRIE, it also undercuts the silicon in the device layer by about 2.7 µm. As a result, silicon structures in the device layer with width less than 5 µm, including the beam, the movable electrode, the springs and the movable combs, are suspended. For the fixed combs, the fingers are suspended but they are connected to a platform anchored to the substrate through the oxide that remains. The fixed combs therefore remain stationary when $V_{comb}$ is applied. Finally, using a shadow mask, we evaporate 5 nm-thick chromium as the adhesion layer and 150 nm-thick gold onto the bonding pads for electric contacts.

The movable combs and the electrode are connected together, supported by four serpentine springs with their other ends anchored to the substrate (Figs. 1a, 1b and 1e). Each comb finger measures 8.3 µm by 2.0 µm by 2.65 µm. The two sets of comb fingers are interdigitated, with a separation of 1.0 µm between adjacent units. Each spring consists of one serpentine turn at a distance of 80 μm from the anchor. The cross-sectional width of 0.78 μm is chosen to be smaller than the thickness of 2.65 μm to minimize motion perpendicular to the substrate in the z direction. From the spring dimensions, we estimate[33] total in-plane spring constant $k_{//}$ of ~ 0.74 N m$^{-1}$ and out-of-plane spring constant $k_\perp$ of ~ 8 N m$^{-1}$. To control the distance $d$ between the movable electrode and the beam from 1.92 μm to 0.26 μm, the electrostatic force produced by the comb drives $F_{comb}$ reaches ~ 1 μN to balance the restoring force from the springs. Such value of $F_{comb}$ exceeds the electrostatic and Casimir forces between the beam and the movable comb by more than a factor of $10^4$. As a result, $d$ is set solely by $V_{comb}$ according to Eq. (2).

2. Sample Characterization

The initial distance $d_0$ is measured from the top view of the beam and the movable electrode (Supplementary Fig. S2a) taken with a Raith 150 nanolithography system in the imaging mode. This machine is capable of producing a beam size of 2 nm according to specifications. Supplementary Figure S2b plots a typical intensity profile along the dotted line in Supplementary Fig. S2a. Within the gap, the intensity is minimum. At the edges, the intensity is maximum. The initial distance between the beam and the movable electrode is determined by counting the number of pixels between the two sharp rising edges. In Supplementary Fig. S2b, the transition region at each edge covers about 3 pixels (each pixel covers 5 nm). After trying various schemes for locating the edge of the beam, we estimate the uncertainty of $d_0$ to be 15 nm.

Supplementary Figure S3a shows a cross sectional scanning electron micrograph of the side-wall profile of a structure (sample B) from the same SOI wafer that is fabricated together with sample A in which the Casimir force was measured. The nominal widths of the beam and the electrode in sample B are identical to sample A. To facilitate cleaving, the length of the structures in sample B is increased to 1.5 mm. The slightly convex slope near the top of the cross section arises due to the periphery of the aluminum etch mask being consumed in the dry etching process. For the rest of the sidewall, the DRIE etching recipe produces a concave profile at an angle of ~ 88 $^\circ$ to the top surface. Supplementary Figure S3b represents the geometry of the beam and the movable electrode that was used in all of our calculations, including the finite element modeling of the electrostatic force and the numerical calculation of the exact Casimir force. The distance $d$ between the beam and the movable electrode is defined as the smallest distance between their opposing sidewalls. The initial distance $d_0$ at $V_{comb} = 0$ is determined from the scanning electron micrographs of the top view of the device (sample A) in which the Casimir force was measured.

3. Comb drive actuator

To control $d$, a comb drive actuator is used to produce a displacement of the movable electrode in the y direction that is proportional to $V_{comb}^2$. Supplementary movies S1 and S2 show the motion of sample C with the comb drives designed to be almost identical to sample A in which the Casimir force was measured. The main difference from sample A is that the width of the movable electrode is reduced to 1 μm. As $V_{comb}$ is increased from 0 V to 9 V, the overlap of the movable and fixed comb fingers increases

(Supplementary movie S1), leading to a decrease in the separation between the beam and the movable electrode (Supplementary movie S2). Through numerical simulations (using a commercial software COMSOL), we verify that the linear dependence of the displacement on $V_{comb}^2$ is better than 1% over the range in our experiment. The slight deviation from linearity changes the value of $d$ by a negligible amount compared to the error bar.

In the absence of the substrate, the electrostatic force generated by the comb drive is entirely in the y direction. With the substrate located 2.0 µm below the comb drive, the electric fields are no longer symmetric on the top and bottom of the comb. It is well-known that when a comb drive is positioned above a conducting substrate, the electrostatic force contains a component normal to the substrate in addition to the lateral component[34]. In our structure, motion of the comb drive perpendicular to the substrate is minimized by two factors. First, the thickness of the springs is designed to be larger than their width by a factor of 3.4, resulting in $k_\perp$ exceeding $k_{//}$ by a factor of ~ 11. Second, the doping in the handle wafer is much lower than the device layer. While the device layer remains conducting at 4 K, the handle wafer is effectively insulating. We calculated that the z displacement of the movable electrode is ~ 12 nm at the largest applied $V_{comb}$. Given that the thickness of the structures is 2.65 µm, the Casimir or electrostatic force is expected to be modified by 0.5% compared to the case when the z displacement is zero. This value is negligible compared to the experimental uncertainty.

In the measurement of the Casimir force between the beam and the movable electrode, electrostatic forces must be minimized. Since the movable electrode and the movable combs are electrically connected, it is sufficient to perform the conventional procedure of applying a voltage $V_e$ to balance the residual potential $V_0$ between these two structures and the beam. The fixed combs, on the other hand, are maintained at $V_{comb}$ up to ~ 12.6 V relative to the movable combs. To prevent the electric fields from the fixed combs from affecting the beam, additional silicon structures are placed between the movable electrode and the movable combs to screen the beam from these fields. Furthermore, we designed and fabricated another device (sample D) to verify that $V_{comb}$ does not lead to any measurable force gradient on the beam. As shown in Supplementary Fig. S4b, sample D is almost identical to the one on which the Casimir force is measured, with the beam and the fixed combs in the usual locations. The only difference is that the other set of combs and the electrode are also immobile due to the absence of etch holes for HF to undercut the silicon device layer. In other words, the only movable structure on sample D is the beam. The distance $d$ between the beam and the electrode remains constant at about 2 μm regardless of $V_{comb}$. When we apply $V_{comb}$ up to 15 V between the two sets of combs, no shifts in the resonant frequency of the beam can be resolved. Therefore we conclude that $V_{comb}$ does not directly exert any measurable electrostatic force on the beam.

**Supplementary References**